\documentclass[letterpaper, 10 pt, conference]{ieeeconf}  %

\IEEEoverridecommandlockouts                              %

\usepackage{graphics} %
\usepackage{epsfig} %
\usepackage{mathptmx} %
\usepackage{times} %
\usepackage{amsmath} %
\usepackage{amssymb}  %
\usepackage{graphicx}
\usepackage{amsmath}
\usepackage{amssymb}
\usepackage{commath}
\usepackage[hidelinks]{hyperref}
\usepackage{tikz}
\usepackage{pgfplots}
\usepackage{siunitx}
\usepackage{multirow}
\usepackage{multirow}
\usepackage{textcomp}
\usepackage{array}
\let\labelindent\relax
\usepackage{enumitem}
\usepackage{setspace}
\usepackage{multirow, makecell}

\usepackage[firstpage=true,color=black,placement=top,scale=1,opacity=1,vshift=-0.3cm]{background}
\SetBgContents{
 \begin{minipage}{1.9\linewidth}
 \small
\textcopyright 2022 IEEE.  Personal use of this material is permitted.  Permission from IEEE must be obtained for all other uses, in any current or future media, including reprinting/republishing this material for advertising or promotional purposes, creating new collective works, for resale or redistribution to servers or lists, or reuse of any copyrighted component of this work in other works. A definitive version was published in \textbf{44th International Engineering in Medicine and Biology Conference (EMBC)} and is available at \url{https://dx.doi.org/10.1109/EMBC48229.2022.9871325}
 \end{minipage}
 }
\title{\LARGE \bf
Automated Characterization of Catalytically Active Inclusion Body Production in Biotechnological Screening Systems*
}

\author{Karina Ruzaeva$^{1,2}$, Kira Küsters$^{2,4}$, Wolfgang Wiechert$^{2,3}$, Benjamin Berkels$^{1}$, \\Marco Oldiges$^{2,4}$
and Katharina Nöh$^{2}$%
\thanks{*This work was performed as part of the Helmholtz School for Data Science in Life, Earth and Energy (HDS-LEE) and received funding from the Helmholtz Association of GRC.}%
\thanks{$^{1}$Aachen Institute for Advanced Study in Computational Engineering Science (AICES), RWTH Aachen University, Aachen, Germany
        {\tt\small ruzaeva@aices.rwth-aachen.de}}%
\thanks{$^{2}$Institute of Bio- and Geosciences, IBG-1: Biotechnology, Forschungszentrum Jülich GmbH, Jülich, Germany
        {\tt\small k.kuesters@fz-juelich.de}}%
\thanks{$^{3}$Institute of Biotechnology, RWTH Aachen University, Aachen, Germany}%
\thanks{$^{4}$Computational Systems Biotechnology (AVT.CSB), RWTH Aachen University, Aachen, Germany}
}

\begin{document}

\maketitle
\thispagestyle{empty}
\pagestyle{empty}

\begin{abstract}
We here propose an automated pipeline for the microscopy image-based characterization of catalytically active inclusion bodies (CatIBs), which includes a fully automatic experimental high-throughput workflow combined with a hybrid approach for multi-object microbial cell segmentation.
For automated microscopy, a CatIB producer strain was cultivated in a microbioreactor from which samples were injected into a flow chamber. The flow chamber was fixed under a microscope and an integrated camera took a series of images per sample.
To explore heterogeneity of CatIB development during the cultivation and track the size and quantity of CatIBs over time, a hybrid image processing pipeline approach was developed, which combines an ML-based detection of in-focus cells with model-based segmentation.
The experimental setup in combination with an automated image analysis unlocks high-throughput screening of CatIB production, saving time and resources.
\newline

\indent \textit{Biotechnological relevance}—  CatIBs have wide application in synthetic chemistry and biocatalysis, but also could have future biomedical applications such as therapeutics. The proposed hybrid automatic image processing pipeline can be adjusted to treat comparable biological microorganisms, where fully data-driven ML-based segmentation approaches are not feasible due to the lack of training data. Our work is the first step towards image-based bioprocess control. %

\end{abstract}

\section{INTRODUCTION}

Inclusion bodies are misfolded proteins, typically regarded as cellular waste products, that occur during overproduction in bacteria such as \textit{Escherichia coli} (\textit{E.~coli}). However, if these inclusion bodies are catalytically active (CatIBs), they become interesting targets for synthetic chemistry and industrial biotechnology, being easy to produce and harvest \cite{Kloss2018, Jaeger2020}. Although CatIB production is compelling, little is known about efficient process conditions. Nowadays, microbioreactors, such as the so-called Biolector system, are used to screen for beneficial (cost-efficient) process conditions \cite{Kuesters2021}. 
For automation, the Biolector is integrated with a liquid handling system, enabling injection of samples into a flow chamber, which is fixed under a microscope. With an integrated camera, a series of images is taken of the samples. 
On the images, CatIBs appear as bright spots within the cells. The number of CatIBs per cell, and time point of the sample are thus detectable analysis parameters that are then correlated with the amount of the protein produced.
This setup allows large-scale optical condition screening to characterize the production of CatIBs. 

\begin{figure}[htp]
  \centering
    \includegraphics[width=\columnwidth]{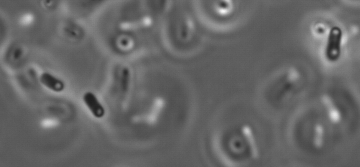}
  \caption{Magnified part of the typical microscopy image showing three in-focus cells (dark ``rods'') with CatIBs and numerous out-of-focus cells.}
    \label{zoomed}
    \vspace{-2ex}
\end{figure}

Due to the large amount of data generated per experiment (500 images per sample, and three samples per one time point), manual analysis of CatIB counts is infeasible. Application of off-the-shelf machine learning (ML)-based segmentation approaches is hampered by a lack of training data. Ground truth (GT), however, is extremely time-intense to produce: since the height of the flow chamber is far larger than the cells, a large fraction of cells is out-of-focus (Figure \ref{zoomed}). Therefore, manual annotation (drawing cell and CatIB outlines) is extremely hard even for experts and risks introducing a bias. 

To objectively analyze the formation of CatIBs in large-scale screens, we propose a hybrid image processing pipeline (HIPP), designed to segment in-focus cells and determine CatIB parameters. With this, analyzing CatIB production at single-cell heterogeneity becomes possible, such as the number and the ratio of the cells carrying CatIBs, along with the average numbers, such as total CatIB and cell areas to monitor the cultivation process. We tested our HIPP with a real data set, showing biotechnologically comprehensible results.

\section{Experimental set-up}

\subsection{Strains, media and cultivation conditions}

\textit{E. coli} DH5 was used as a cloning host for the generation of plasmid pET28a::\textit{Ec}LDCc::SG::TDoT. Heterologous expression of the gene fusions was performed using \textit{E.~coli} BL21(DE3). Lysogeny broth (LB) medium was employed for the cultivation of strains during the cloning procedure and for the precultures of the expression strains. %
For automated microscopy experiments, the expression strains were cultivated in 48-well FlowerPlates (m2p-labs GmbH, Baesweiler, Germany) in M9-AI medium \cite{Kuesters2021} in a BioLector (m2p-labs GmbH, Baesweiler, Germany). %
After precultivation, expression cultures were inoculated at OD600, and the cells were cultivated at 37°C for 3~hours at 1000~rpm. Then the temperature was decreased to 15°C and the expression ensued for 69~hours under the same shaking conditions. 
All cultures were supplemented with 50~µg/ml kanamycin for plasmid maintenance.

\subsection{Automated experimentation and microscopy}

 CatIB strains were cultivated in a BioLector, which was integrated into a liquid handling system (Freedom Evo200, Tecan, Männedorf, Switzerland). Every 4.5 hours, cells were harvested from each well of the FlowerPlate (the last three wells were harvested at 72~hours) and stored in a deep well plate at 4°C on the robotic deck until automated microscopy was performed. For sample injection, a self-built injection station \cite{Jansen2021} was used. The station was connected via a tubing (VWR, Darmstadt, Germany) to a flow chamber (height: 20~µm, length: 58.5~mm, width: 800~µm, microfluidic ChipShop, Jena, Germany). The chamber was fixed on an inverted Nikon Eclipse Ti microscope (Nikon GmbH, Düsseldorf, Germany) equipped with a CFI Plan Apo Lambda 100x Oil objective (Nikon GmbH, Düsseldorf, Germany). 300~µL of each sample, followed by 400~µL H\textsubscript{2}O were injected in the injection station. After the injection of 600~µL the flow was set to zero to allow for image acquisition. A 1~µL pulse with a velocity of 1~µL/switch after a 4 min delay was performed three times to flush in new cells. 500 images (one image per second) were taken of each sample in the flow chamber, resulting in 1500 images for each time point. Images were taken with a Thorlab camera DCC154M-GL (Thorlabs Inc., Newton, New Jersey, USA). 

\section{Hybrid image processing pipeline}

The goal of the proposed HIPP is to segment the in-focus cells and their CatIBs (if present) separately and afterwards, classify the segmented cells, based on the number of the CatIBs.
To accomplish the goal, considering the large fraction of free-flowing out-of-focus cells, we split multi-object (in-focus cells) and multi-label segmentation (number of CatIBs in a cell) into two major steps: an ML-based single-class detection approach and a model-based segmentation step, which includes variational-based and thresholding-based segmentation. The introduced separation has four reasons:
\begin{enumerate}[wide, labelindent=5pt]
    \item Despite the availability of the multi-object segmentation methods, like \cite{He2017}, their main limitation is the lack of training data, being, in our case, very laborious to produce. Here, unlike precise pixel-wise image annotation that is necessary for the training of a multi-object pipeline, we use an easy to create single-class training dataset, since only bounding boxes are needed to train the detection framework.
    \item Although, the detection framework we here use supports multi-class detection, and could be used as a classifier for the different types of cells (none, 1 or 2 CatIBs per cell), it still requires a bigger training dataset, and due to the small feature (CatIB) size, compared to the image size, adequate classification accuracy was not reached in our preliminary numerical experiments.
    \item Another important benefit of the separation of detection and the segmentation step is simplifying the problem by reformulating multi-object segmentation as many single-object segmentation tasks. This unlocks single-object model-based segmentation methods, such as variational-based segmentation and thresholding.
    \item One more advantage of the splitting is that the segmentation can be done in parallel for the cells since each detected bounding box contains only one cell, which can be processed independently.
\end{enumerate}

\subsection{Detection} 
\label{secdet}

To detect the in-focus cells, the YOLOv5 detection framework, one of the fastest available detection frameworks \cite{Yolov5}, was used. However, using another detection framework is certainly possible.
The training data was generated by manual annotation of four images, where only in-focus cells were labeled, resulting in 151 objects (cells). To artificially increase the training dataset YOLOv5's built-in augmentation was used, which includes image saturation, hue and value augmentation; rotation, translation and scaling; horizontal and vertical flips and image mosaic augmentation.
The detection was performed on 15 manually chosen images with a steady flow for each of the 16~time points. The resulting output is discussed in Section \ref{result}.

\begin{figure}[!h]
  \centering
  \resizebox{.325\columnwidth}{!}{%
  \begin{tikzpicture}

\definecolor{color0}{rgb}{0.75,0.75,0}

\begin{axis}[
    hide axis,
axis equal image,
tick align=outside,
tick pos=left,
x grid style={white!69.0196078431373!black},
xmin=-0.5, xmax=1348.225,
xtick style={color=black},
y dir=reverse,
y grid style={white!69.0196078431373!black},
ymin=-0.5, ymax=1023.5,
ytick style={color=black}
]
\addplot graphics [includegraphics cmd=\pgfimage,xmin=-0.5, xmax=1279.5, ymin=1023.5, ymax=-0.5] {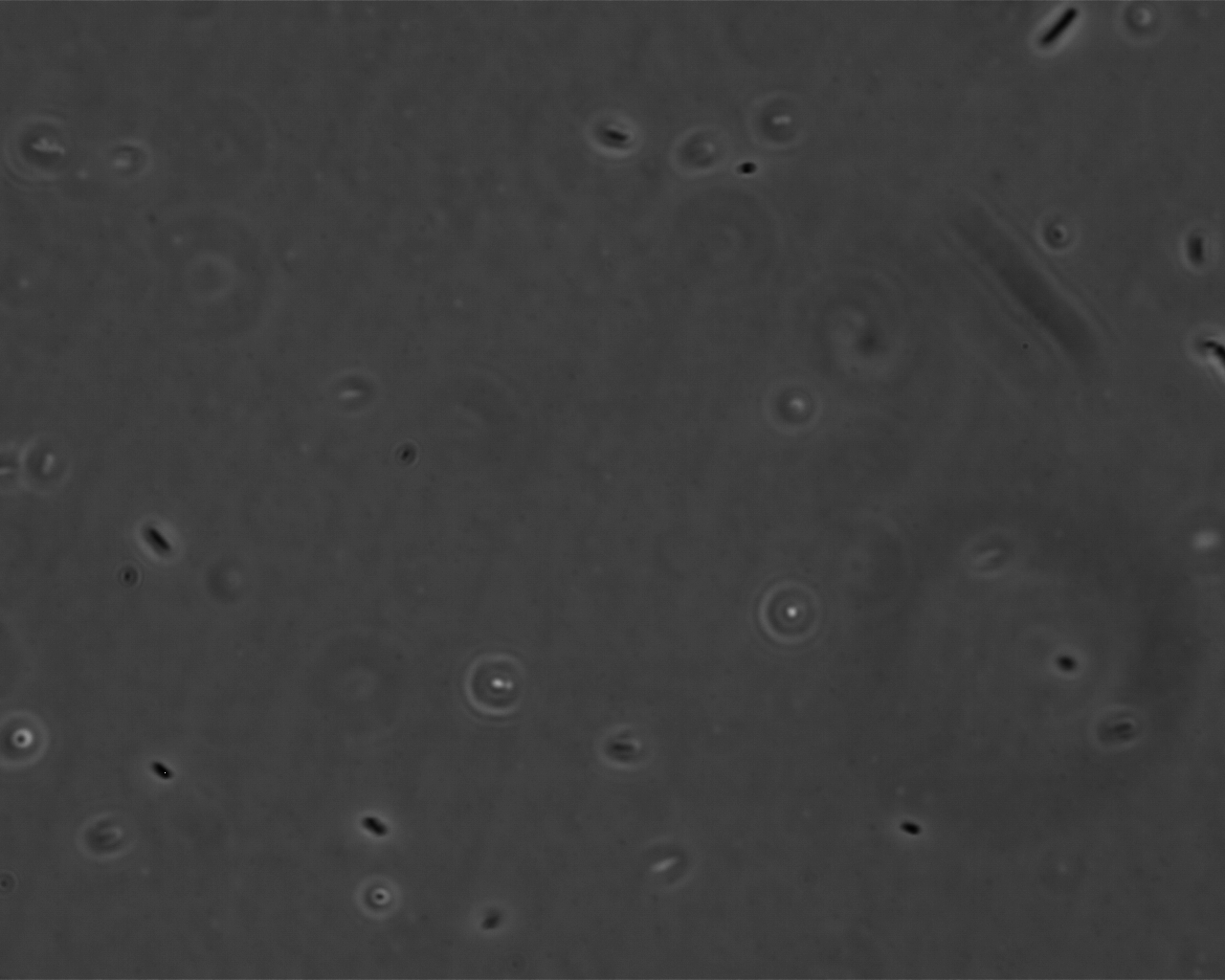};
\addplot [semithick, color0]
table {%
1244 337
1244 401
1279 401
1279 337
1244 337
};
\addplot [semithick, color0]
table {%
611 118
611 158
671 158
671 118
611 118
};
\addplot [semithick, color0]
table {%
929 847
929 880
973 880
973 847
929 847
};
\addplot [semithick, color0]
table {%
761 159
761 188
796 188
796 159
761 159
};
\addplot [semithick, color0]
table {%
143 537
143 589
189 589
189 537
143 537
};
\addplot [semithick, color0]
table {%
365 843
365 881
416 881
416 843
365 843
};
\addplot [semithick, color0]
table {%
148 783
148 824
188 824
188 783
148 783
};
\end{axis}

\end{tikzpicture}}
  \resizebox{.325\columnwidth}{!}{%
  \begin{tikzpicture}

\definecolor{color0}{rgb}{0,1,0}

\begin{axis}[
    hide axis,
axis equal image,
tick align=outside,
tick pos=left,
x grid style={white!69.0196078431373!black},
xmin=-0.5, xmax=1348.225,
xtick style={color=black},
y dir=reverse,
y grid style={white!69.0196078431373!black},
ymin=-0.5, ymax=1023.5,
ytick style={color=black}
]
\addplot graphics [includegraphics cmd=\pgfimage,xmin=-0.5, xmax=1279.5, ymin=1023.5, ymax=-0.5] {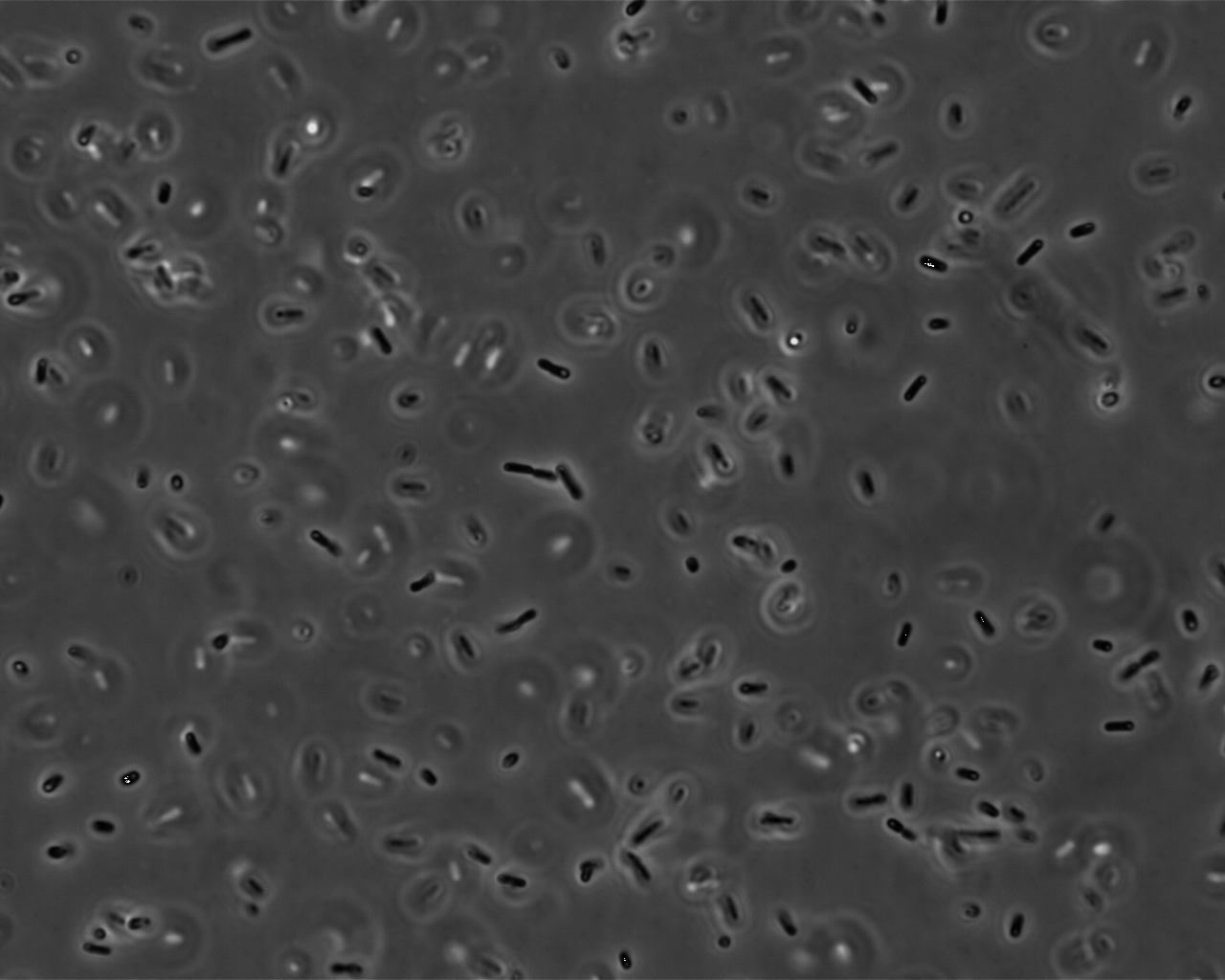};
\addplot [semithick, color0]
table {%
87 846
87 878
129 878
129 846
87 846
};
\addplot [semithick, color0]
table {%
465 650
465 693
508 693
508 650
465 650
};
\addplot [semithick, color0]
table {%
479 870
479 913
521 913
521 870
479 870
};
\addplot [semithick, color0]
table {%
872 811
872 856
941 856
941 811
872 811
};
\addplot [semithick, color0]
table {%
514 472
514 512
586 512
586 472
514 472
};
\addplot [semithick, color0]
table {%
242 906
242 949
291 949
291 906
242 906
};
\addplot [semithick, color0]
table {%
1113 323
1113 383
1177 383
1177 323
1113 323
};
\addplot [semithick, color0]
table {%
705 572
705 603
737 603
737 572
705 572
};
\addplot [semithick, color0]
table {%
646 883
646 932
694 932
694 883
646 883
};
\addplot [semithick, color0]
table {%
1247 684
1247 734
1279 734
1279 684
1247 684
};
\addplot [semithick, color0]
table {%
927 179
927 232
977 232
977 179
927 179
};
\addplot [semithick, color0]
table {%
375 767
375 817
433 817
433 767
375 767
};
\addplot [semithick, color0]
table {%
77 973
77 1006
126 1006
126 973
77 973
};
\addplot [semithick, color0]
table {%
804 572
804 607
838 607
838 572
804 572
};
\addplot [semithick, color0]
table {%
640 982
640 1015
669 1015
669 982
640 982
};
\addplot [semithick, color0]
table {%
1252 379
1252 416
1279 416
1279 379
1252 379
};
\addplot [semithick, color0]
table {%
759 697
759 735
814 735
814 697
759 697
};
\addplot [semithick, color0]
table {%
781 833
781 877
849 877
849 833
781 833
};
\addplot [semithick, color0]
table {%
378 328
378 382
417 382
417 328
378 328
};
\addplot [semithick, color0]
table {%
957 320
957 354
1003 354
1003 320
957 320
};
\addplot [semithick, color0]
table {%
768 294
768 352
824 352
824 294
768 294
};
\addplot [semithick, color0]
table {%
1216 89
1216 134
1256 134
1256 89
1216 89
};
\addplot [semithick, color0]
table {%
882 476
882 537
929 537
929 476
882 476
};
\addplot [semithick, color0]
table {%
982 789
982 824
1034 824
1034 789
982 789
};
\addplot [semithick, color0]
table {%
112 793
112 830
157 830
157 793
112 793
};
\addplot [semithick, color0]
table {%
429 792
429 829
468 829
468 792
429 792
};
\addplot [semithick, color0]
table {%
1132 658
1132 688
1172 688
1172 658
1132 658
};
\addplot [semithick, color0]
table {%
310 536
310 598
370 598
370 536
310 536
};
\addplot [semithick, color0]
table {%
1156 667
1156 723
1228 723
1228 667
1156 667
};
\addplot [semithick, color0]
table {%
422 582
422 629
464 629
464 582
422 582
};
\addplot [semithick, color0]
table {%
514 772
514 812
554 812
554 772
514 772
};
\addplot [semithick, color0]
table {%
508 900
508 940
564 940
564 900
508 900
};
\addplot [semithick, color0]
table {%
149 174
149 223
190 223
190 174
149 174
};
\addplot [semithick, color0]
table {%
916 842
916 888
973 888
973 842
916 842
};
\addplot [semithick, color0]
table {%
1223 623
1223 670
1266 670
1266 623
1223 623
};
\addplot [semithick, color0]
table {%
803 938
803 985
843 985
843 938
803 938
};
\addplot [semithick, color0]
table {%
180 751
180 801
220 801
220 751
180 751
};
\addplot [semithick, color0]
table {%
203 18
203 66
277 66
277 18
203 18
};
\addplot [semithick, color0]
table {%
1054 239
1054 287
1102 287
1102 239
1054 239
};
\addplot [semithick, color0]
table {%
568 471
568 535
624 535
624 471
568 471
};
\addplot [semithick, color0]
table {%
882 70
882 117
926 117
926 70
882 70
};
\addplot [semithick, color0]
table {%
31 793
31 838
77 838
77 793
31 793
};
\addplot [semithick, color0]
table {%
1104 218
1104 258
1157 258
1157 218
1104 218
};
\addplot [semithick, color0]
table {%
951 254
951 297
1004 297
1004 254
951 254
};
\addplot [semithick, color0]
table {%
1042 941
1042 988
1078 988
1078 941
1042 941
};
\addplot [semithick, color0]
table {%
506 623
506 673
574 673
574 623
506 623
};
\addplot [semithick, color0]
table {%
924 639
924 686
961 686
961 639
924 639
};
\addplot [semithick, color0]
table {%
547 362
547 405
605 405
605 362
547 362
};
\addplot [semithick, color0]
table {%
934 378
934 429
979 429
979 378
934 378
};
\addplot [semithick, color0]
table {%
1142 742
1142 772
1194 772
1194 742
1142 742
};
\addplot [semithick, color0]
table {%
1008 625
1008 676
1052 676
1052 625
1008 625
};
\end{axis}

\end{tikzpicture}}
  \resizebox{.325\columnwidth}{!}{%
  \begin{tikzpicture}

\definecolor{color0}{rgb}{1,0,0}

\begin{axis}[
    hide axis,
axis equal image,
tick align=outside,
tick pos=left,
x grid style={white!69.0196078431373!black},
xmin=-0.5, xmax=1348.225,
xtick style={color=black},
y dir=reverse,
y grid style={white!69.0196078431373!black},
ymin=-0.5, ymax=1023.5,
ytick style={color=black}
]
\addplot graphics [includegraphics cmd=\pgfimage,xmin=-0.5, xmax=1279.5, ymin=1023.5, ymax=-0.5] {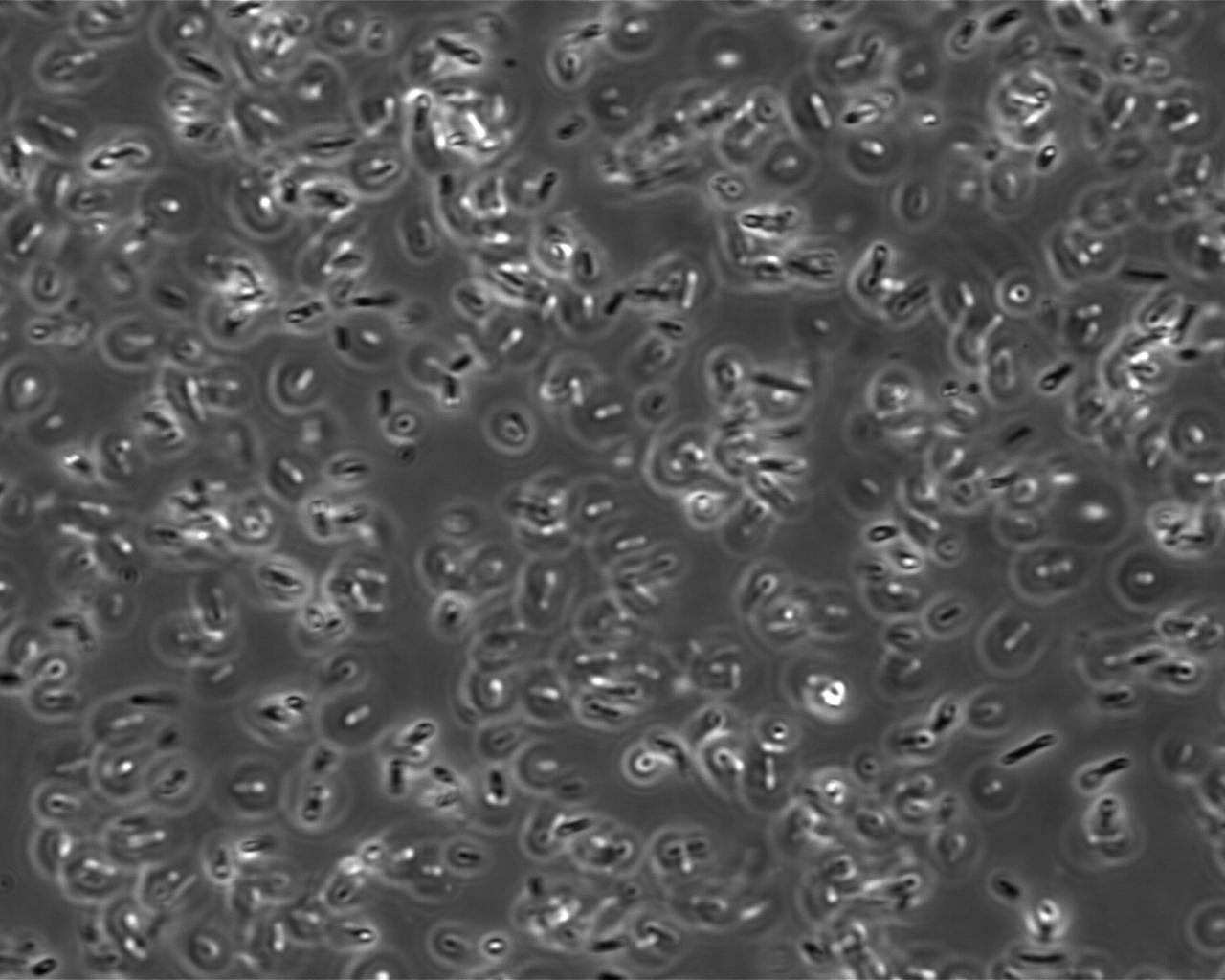};
\addplot [semithick, color0]
table {%
1041 435
1041 475
1089 475
1089 435
1041 435
};
\addplot [semithick, color0]
table {%
356 296
356 333
428 333
428 296
356 296
};
\addplot [semithick, color0]
table {%
458 356
458 397
505 397
505 356
458 356
};
\addplot [semithick, color0]
table {%
550 170
550 227
597 227
597 170
550 170
};
\addplot [semithick, color0]
table {%
285 456
285 532
337 532
337 456
285 456
};
\addplot [semithick, color0]
table {%
1118 778
1118 835
1191 835
1191 778
1118 778
};
\addplot [semithick, color0]
table {%
770 373
770 430
857 430
857 373
770 373
};
\addplot [semithick, color0]
table {%
1030 748
1030 815
1120 815
1120 748
1030 748
};
\addplot [semithick, color0]
table {%
1070 364
1070 424
1141 424
1141 364
1070 364
};
\end{axis}

\end{tikzpicture}}
  
  \vspace{0.5cm}
    \begin{tikzpicture}

\definecolor{color0}{rgb}{0.12156862745098,0.466666666666667,0.705882352941177}
\definecolor{color1}{rgb}{0.75,0.75,0}
\begin{axis}[
tick align=outside,
tick pos=left,
x grid style={white!69.0196078431373!black},
xlabel={time \lbrack h\rbrack},
xmin=0.0249999999999999, xmax=76.475,
xtick style={color=black},
y grid style={white!69.0196078431373!black},
ylabel={N cells},
ymin=0, ymax=710.85,
ytick style={color=black},
yscale=0.5,
]
\draw[draw=none,fill=color1] (axis cs:3.5,0) rectangle (axis cs:5.5,114);
\draw[draw=none,fill=color0] (axis cs:8,0) rectangle (axis cs:10,196);
\draw[draw=none,fill=color0] (axis cs:12.5,0) rectangle (axis cs:14.5,344);
\draw[draw=none,fill=color0] (axis cs:17,0) rectangle (axis cs:19,260);
\draw[draw=none,fill=color0] (axis cs:21.5,0) rectangle (axis cs:23.5,335);
\draw[draw=none,fill=color0] (axis cs:26,0) rectangle (axis cs:28,457);
\draw[draw=none,fill=color0] (axis cs:30.5,0) rectangle (axis cs:32.5,618);
\draw[draw=none,fill=green] (axis cs:35,0) rectangle (axis cs:37,677);
\draw[draw=none,fill=color0] (axis cs:39.5,0) rectangle (axis cs:41.5,431);
\draw[draw=none,fill=color0] (axis cs:44,0) rectangle (axis cs:46,503);
\draw[draw=none,fill=color0] (axis cs:48.5,0) rectangle (axis cs:50.5,628);
\draw[draw=none,fill=color0] (axis cs:53,0) rectangle (axis cs:55,465);
\draw[draw=none,fill=color0] (axis cs:57.5,0) rectangle (axis cs:59.5,639);
\draw[draw=none,fill=color0] (axis cs:62,0) rectangle (axis cs:64,453);
\draw[draw=none,fill=color0] (axis cs:66.5,0) rectangle (axis cs:68.5,441);
\draw[draw=none,fill=red] (axis cs:71,0) rectangle (axis cs:73,234);
\end{axis}

\end{tikzpicture}
    
  \caption{
           Number of detected cells at each time point (bottom) and detection examples for images from different time points (top, marked with the corresponding colors).}
    \label{detection}
\end{figure}
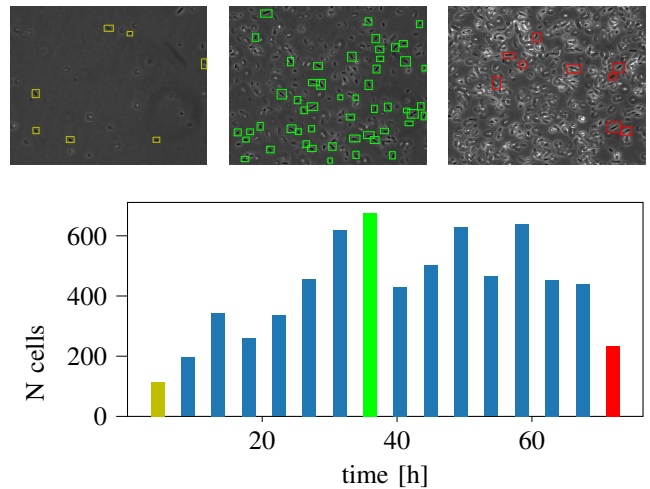

Figure \ref{detection} shows detection examples for images from the beginning (4.5 hours, yellow), middle (36 hours, green), and the end (72 hours, red) of the dataset and the number of the detected cells on 15 images for each time point. In the bar chart, the positive slope in the first part of the experiment (4.5-36 hours) coincides with the natural growth of the cell colonies. The decline of the number of detected cells during the last part (60.5-72 hours) is explained by the fact, that during the last sampling points, \textit{E. coli} cells were too crowded for exact automated image analysis. Especially in the last time point (Figure \ref{detection}), nearly all in-focus cells overlapped with out-of-focus cells, hampering the automated analysis. In the future, to improve the automated microscopy system, a dilution step of the cells should be added at a specific \textit{E. coli} density to ensure correct image analysis over the whole cultivation time. Moreover, the detection accuracy can be improved by using a bigger training dataset, since we use an ML-based detection.

\subsection{Segmentation}

After the detection, each detected cell is segmented individually, by considering the image tile inside the detected bounding box. There, the segmentation is divided into two steps: first, cell-body segmentation (no CatIBs are assumed) and next CatIB(s) segmentation in each cell.
To this end, we use a model-based segmentation approach on each single-cell image tile. Direct cell body segmentation is complicated by the fact that CatIBs are as bright as the background, or even brighter, and in most cases missegmented as background. To avoid this problem, we ``fill'' the CatIBs using a morphological closing operation \cite{RafaelC.Gonzalez2018}, which combines the application of the dilation (minimum filter, as the cell is darker than the background) and erosion (maximum filter) to the dilated image tile. Then, to smooth the artifacts resulting from the closing operation and to avoid local minima problems during the following variational segmentation step, Gaussian blur is applied to the obtained ``closed'' (inclusion-less) image tile.

To separate the main (fully present) cell from parts of cells in the neighborhood included in the corresponding bounding box, a variational B-splines-based segmentation \cite{Ruzaeva2022} is performed, where the cell shape is modeled as a straight rod (by fixing the curvature parameters $d$ and $e$ in the notation of \cite{Ruzaeva2022}, to zero), which consists of six control points. As the shape of the cell is more flexible than the shape model used in the variational segmentation step, to obtain the fine cell contour, global manually set thresholding (the same threshold value was applied to the entire dataset) is applied to the pixels which belong to the interior of the spline, which results in a binary pixel mask of the cell body. The suggested spline fit is necessary if parts of other cells are in the bounding box, but not necessary if sparse cell flow can be guaranteed (i.e. with dilution). To acquire the mask for the CatIB(s) of the target cell, thresholding was also applied to the original image tile, where the resulting CatIB mask is multiplied by the cell body mask, obtained at the previous step. 

Since every cell has a bright surrounding “halo”, which can be missegmented as CatIB, every CatIB mask instance is checked for its area and its eccentricity. The CatIB mask components with high eccentricity (more than 80\%) and small area (2 or fewer pixels) are assumed to be outliers and removed from the mask. Since the presence of CatIBs can not be guaranteed and the number of the labels is thus unknown, a multi-label histogram-based threshold cannot be applied.

\begin{figure}[!ht]
  \centering
  \resizebox{!}{2.75cm}{%
  \begin{tikzpicture}

\definecolor{color0}{rgb}{0.75,0.75,0}

\begin{axis}[
    hide axis,
axis equal image,
tick align=outside,
tick pos=left,
x grid style={white!69.0196078431373!black},
xmin=-0.5, xmax=41.5,
xtick style={color=black},
y dir=reverse,
y grid style={white!69.0196078431373!black},
ymin=-0.5, ymax=54.5,
ytick style={color=black}
]
\addplot graphics [includegraphics cmd=\pgfimage,xmin=-0.5, xmax=41.5, ymin=54.5, ymax=-0.5] {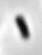};
\node [scale=2, above, black] at (40,95) {a)};
\addplot [ultra thick, black]
table {%
28 51
38 51
};
\node [above, black] at (330,510) {\SI{1}{\micro\metre}};
\addplot [line width=1.5pt, green]
table {%
23 39
24 39
25 39
26 40
27 39
28 39
29 38
29 37
30 36
30 35
30 34
30 33
29 32
29 31
29 30
29 29
28 28
28 27
27 26
27 25
27 24
27 23
26 22
26 21
25 20
25 19
24 18
24 17
23 16
22 16
21 15
20 15
19 15
18 16
17 16
16 17
16 18
16 19
16 20
16 21
16 22
16 23
16 24
16 25
17 26
17 27
18 28
18 29
19 30
19 31
20 32
20 33
20 34
21 35
21 36
21 37
22 38
23 39
};
\end{axis}

\end{tikzpicture}}
  \resizebox{!}{2.75cm}{%
  \begin{tikzpicture}

\definecolor{color0}{rgb}{0.75,0.75,0}

\begin{axis}[
    hide axis,
axis equal image,
tick align=outside,
tick pos=left,
x grid style={white!69.0196078431373!black},
xmin=-0.5, xmax=53.5,
xtick style={color=black},
y dir=reverse,
y grid style={white!69.0196078431373!black},
ymin=-0.5, ymax=42.5,
ytick style={color=black}
]
\addplot graphics [includegraphics cmd=\pgfimage,xmin=-0.5, xmax=53.5, ymin=42.5, ymax=-0.5] {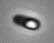};
\node [scale=2, above, black] at (40,80) {b)};
\addplot [ultra thick, black]
table {%
40 40
50 40
};
\node [above, black] at (450,400) {\SI{1}{\micro\metre}};
\addplot [ultra thick, red]
table {%
31 28
32 28
33 28
34 28
35 28
36 27
37 26
37 25
37 24
36 23
35 22
34 21
33 20
32 20
31 20
30 20
29 20
28 20
27 21
26 22
25 23
25 24
25 25
26 26
27 27
28 27
29 28
30 28
31 28
};
\addplot [line width=1.5pt, green]
table {%
27 28
28 29
29 29
30 29
31 29
32 30
33 30
34 30
35 30
36 29
37 29
38 29
39 28
39 27
40 26
40 25
40 24
40 23
39 22
39 21
38 20
37 19
36 19
35 19
34 18
33 18
32 18
31 18
30 18
29 17
28 17
27 17
26 17
25 16
24 16
23 16
22 15
21 15
20 15
19 15
18 15
17 15
16 15
15 16
14 16
13 17
13 18
13 19
13 20
13 21
13 22
14 23
15 24
16 24
17 25
18 25
19 26
20 26
21 26
22 27
23 27
24 27
25 28
26 28
27 28
};
\end{axis}

\end{tikzpicture}}
  \resizebox{!}{2.75cm}{%
  \begin{tikzpicture}

\definecolor{color0}{rgb}{0.75,0.75,0}

\begin{axis}[
    hide axis,
axis equal image,
tick align=outside,
tick pos=left,
x grid style={white!69.0196078431373!black},
xmin=-0.5, xmax=55.5,
xtick style={color=black},
y dir=reverse,
y grid style={white!69.0196078431373!black},
ymin=-0.5, ymax=59.5,
ytick style={color=black}
]
\addplot graphics [includegraphics cmd=\pgfimage,xmin=-0.5, xmax=55.5, ymin=59.5, ymax=-0.5] {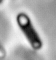};
\node [scale=2, above, black] at (40,95) {c)};
\addplot [ultra thick, black]
table {%
42 56
52 56
};
\node [above, black] at (470,560) {\SI{1}{\micro\metre}};
\addplot [line width=1.5pt, green]
table {%
30 44
31 45
32 46
33 46
34 47
35 47
36 47
37 47
38 47
39 46
40 46
41 45
41 44
41 43
41 42
41 41
41 40
41 39
40 38
40 37
40 36
39 35
39 34
38 33
38 32
37 31
36 30
36 29
35 28
34 27
34 26
33 25
33 24
32 23
32 22
31 21
30 20
30 19
29 18
28 16
28 17
27 15
26 14
25 14
24 13
23 13
22 13
21 13
20 13
19 14
18 15
17 16
17 17
17 18
17 19
17 20
17 21
17 22
18 23
18 24
18 25
19 26
19 27
20 28
21 29
21 30
22 31
23 32
23 33
24 34
25 35
25 36
26 37
27 38
27 39
28 40
28 41
29 42
29 43
30 44
};

\addplot [thick, red]
table {%
23 25
24 25
25 24
26 23
27 22
27 21
27 20
27 19
27 18
26 17
25 16
24 15
23 15
22 15
21 15
20 15
19 16
19 17
18 18
18 19
18 20
19 21
19 22
19 23
20 24
21 25
22 25
23 25
};
\addplot [thick, red]
table {%
36 47
37 46
38 45
38 44
38 43
37 42
36 42
35 42
34 42
33 43
33 44
33 45
34 46
35 47
36 47
};
\end{axis}
\end{tikzpicture}}
  
  \caption{\label{fig:cell}
           Example segmentation results for the three single-cell image tiles, selected from the results, which illustrate different cell classes (0 (a), 1 (b), and 2 (c) CatIBs) from the validation dataset. The green contour represents the cell body, the red contour – the CatIBs (if present).}
\end{figure}
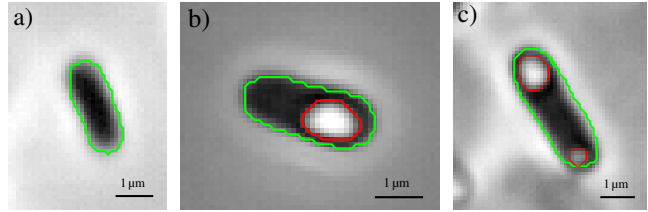

\section{Results}
\label{result}

\subsection{Accuracy quantification}

Since no benchmark datasets are available for our application setting, we created GT data ourselves. Therefore, the validation dataset is rather small. As a validation dataset for the segmentation, 30 single-cell image tiles, obtained at the detection step, were used, where manual pixel-wise segmentation annotation was performed by an expert. The dataset contains ten image tiles of each class from different time points, where for the ``1 CatIB'' and the ``2 CatIBs'' classes, the GT CatIB masks are provided as well.

As a quality metric for the proposed segmentation algorithm, we used the average Dice score for the cell and for the CatIBs:
\begin{equation}
\mathrm{Dice}(\hat{I_i},(I_i)_{i=1}^N)=\frac{1}{N}\sum_{i=1}^{N}\frac{2|\hat{I_i}\cap I_i|}{|\hat{I_i}|+ |I_i|}
\label{eq}
\end{equation}
Here, $I_i$ is the binary mask (either cell body or CatIB(s)), obtained with the proposed pipeline; $\hat{I_i}$ is the GT binary mask of the respective object, from the manual annotation by the expert and $N=30$ is the number of images tiles in the validation dataset.

In the case of the first class, with no CatIBs present and no CatIBs predicted, the Dice score for the frame is set to be 1.
The resulting quality metrics for the empirically tuned parameter set (see below), which provides the best scores, are $\mathrm{Dice}_\text{cell}=0.88$, $\mathrm{Dice}_\text{CatIB}=0.78$. The parameter set includes the cell body and the CatIB thresholds; sizes of the masks for the dilation and erosion, and the weights for the variational spline-based segmentation step \cite{Ruzaeva2022}.

The low $\mathrm{Dice}_\text{CatIB}$ score can be explained by the fact that generally, the CatIBs have around 3-10 pixels area and even the slightest (1 pixel) deviation of the result from the GT affects the score strongly.

The classification of the cells was performed, based on the segmentation result, where cells with no CatIBs segmented belong to the first class, cells with one solid CatIB (after the area and eccentricity checks) belong to the second class, and cells with two separate (not 8-connected) CatIBs detected, belong to the last class. The confusion matrix is shown below (Table~\ref{cm}).
 
\begin{table}[!ht]
    \centering
    \caption{ Confusion matrix for segmentation-based cells classification (3 classes: 0, 1, 2 CatIB(s) per cell)}
    \label{cm}
    {\renewcommand{\arraystretch}{1.2}
\begin{tabular}{@{}cc|ccc@{}}
\multicolumn{1}{c}{} &\multicolumn{1}{c}{} &\multicolumn{3}{c}{Predicted} \\ 
\multicolumn{1}{c}{} & 
\multicolumn{1}{c|}{} & 
\multicolumn{1}{c}{No CatIB} & 
\multicolumn{1}{c}{1 CatIB} &
\multicolumn{1}{c}{2 CatIBs} \\ 
\cline{2-5}
\multirow[c]{3}{*}{\rotatebox[origin=tr]{90}{Actual}}
& No CatIB  & \textbf{10} & 0 &0  \\
& 1 CatIB  & 1   & \textbf{9}&0 \\
& 2 CatIBs  & 1   & 4&\textbf{5} \\ 
\cline{2-5}
\end{tabular}}
\end{table}

\subsection{CatIB characterization}

To evaluate the automatic analysis, we selected and manually counted the cells with CatIB(s) and the total number of cells for each time point on five images to calculate the fraction of the cells with CatIB(s). The GT results were compared against the results, obtained with the proposed HIPP on 15 images for each time point. 

\begin{figure}[!ht]
  \centering
  \resizebox{.9\columnwidth}{!}{%
  \begin{tikzpicture}

\begin{axis}[
tick align=outside,
tick pos=left,
x grid style={white!69.0196078431373!black},
xlabel={time\lbrack h\rbrack},
xmin=1.125, xmax=75.375,
xtick style={color=black},
y grid style={white!69.0196078431373!black},
ylabel={N cells with IB(s) / N cells},
ymin=-0.0416839485690585, ymax=0.875362919950228,
ytick style={color=black},
ytick={-0.2,0,0.2,0.4,0.6,0.8,1},
yticklabels={\ensuremath{-}0.2,0.0,0.2,0.4,0.6,0.8,1.0},
y post scale=0.5,
]
\addplot [semithick, red]
table {%
4.5 0.043859649122807
9 0.0459183673469388
13.5 0.0465116279069767
18 0.107692307692308
22.5 0.0477611940298507
27 0.0590809628008753
31.5 0.0760517799352751
36 0.0915805022156573
40.5 0.327146171693735
45 0.443339960238569
49.5 0.504777070063694
54 0.395698924731183
58.5 0.561815336463224
63 0.593818984547461
67.5 0.589569160997732
72 0.491452991452991
};
\addplot [semithick, blue]
table {%
4.5 0
9 0
13.5 0
18 0
22.5 0
27 0.0443458980044346
31.5 0.170490423068828
36 0.241154726280584
40.5 0.478248343416883
45 0.541564097708767
49.5 0.721533258173619
54 0.640257945647167
58.5 0.806080147397513
63 0.83031035872632
67.5 0.83367897138117
72 0.771734077173408
};
\end{axis}

\end{tikzpicture}}
  \caption{The fraction of the CatIB-carrier cells over time during the cultivation experiment, obtained by manual counting (blue) and the proposed HIPP (red).}
    \label{gtvsauro}
\end{figure}
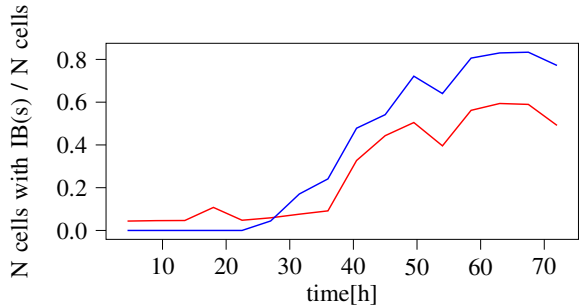

In Figure \ref{gtvsauro}, in the second half of the experiment, an underestimation of cells with CatIBs is observed, which correlates with the classification results (Section~\ref{cm}).
Nevertheless, the obtained automatic result is qualitatively matching the manual result (with Pearson correlation coefficient = 98~\%) and is useful for experts to adjust the experimental design, conditions control, etc. 

\begin{figure}[!ht]
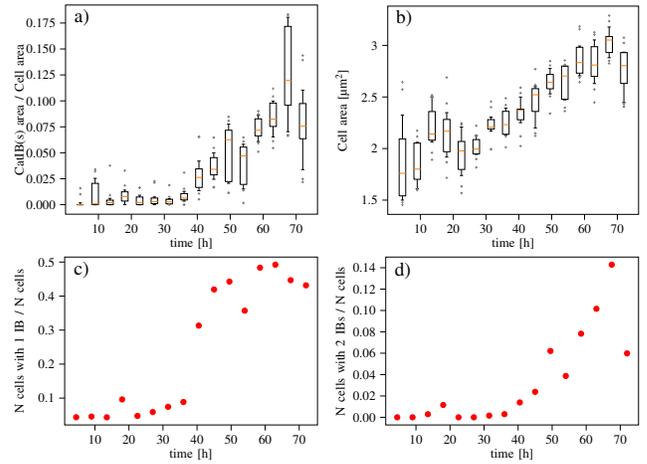

  \centering
  \resizebox{.48\columnwidth}{!}{%
  \input{figure/sinc_over_scells}}
  \resizebox{.48\columnwidth}{!}{%
  \input{figure/scells}}
  \resizebox{.48\columnwidth}{!}{%
  \begin{tikzpicture}

\begin{axis}[
tick align=outside,
tick pos=left,
x grid style={white!69.0196078431373!black},
xlabel={time \lbrack h\rbrack},
xmin=1.125, xmax=75.375,
xtick style={color=black},
y grid style={white!69.0196078431373!black},
ylabel={N cells with 1 IB / N cells},
ymin=0.0211711971867139, ymax=0.514707184660404,
ytick style={color=black},
ytick={0,0.1,0.2,0.3,0.4,0.5,0.6},
yticklabels={0.0,0.1,0.2,0.3,0.4,0.5,0.6},
y post scale=0.8,
]
\node [scale=1.5, above, black] at (50,410) {c)};
\addplot [semithick, red, mark=*, mark size=2, mark options={solid}, only marks,x=0.5cm,y=1cm]
table {%
4.5 0.043859649122807
9 0.0459183673469388
13.5 0.0436046511627907
18 0.0961538461538462
22.5 0.0477611940298507
27 0.0590809628008753
31.5 0.0744336569579288
36 0.0886262924667651
40.5 0.31322505800464
45 0.41948310139165
49.5 0.442675159235669
54 0.356989247311828
58.5 0.483568075117371
63 0.492273730684327
67.5 0.44671201814059
72 0.431623931623932
};
\end{axis}

\end{tikzpicture}}
  \resizebox{.48\columnwidth}{!}{%
  \begin{tikzpicture}

\begin{axis}[
tick align=outside,
tick pos=left,
x grid style={white!69.0196078431373!black},
xlabel={time \lbrack h\rbrack},
xmin=1.125, xmax=75.375,
xtick style={color=black},
y grid style={white!69.0196078431373!black},
ylabel={N cells with 2 IBs / N cells},
ymin=-0.00714285714285714, ymax=0.15,
ytick style={color=black},
ytick={-0.02,0,0.02,0.04,0.06,0.08,0.1,0.12,0.14,0.16},
yticklabels={\ensuremath{-}0.02,0.00,0.02,0.04,0.06,0.08,0.10,0.12,0.14,0.16},
y post scale=0.8,
]
\node [scale=1.5, above, black] at (50,130) {d)};
\addplot [semithick, red, mark=*, mark size=2, mark options={solid}, only marks]
table {%
4.5 0
9 0
13.5 0.00290697674418605
18 0.0115384615384615
22.5 0
27 0
31.5 0.00161812297734628
36 0.00295420974889217
40.5 0.0139211136890951
45 0.0238568588469185
49.5 0.0621019108280255
54 0.0387096774193548
58.5 0.0782472613458529
63 0.101545253863135
67.5 0.142857142857143
72 0.0598290598290598
};
\end{axis}

\end{tikzpicture}}
  
  \caption{CatIB development over time: (a) CatIB area fraction and (b) Average cell area, with the median (orange), the 25/75~\% quartiles (black box) with 10~\%/ 90~\% whiskers and the outliers (gray crosses); (c) Fraction of the cells with one CatIB; (d) Fraction of cells with two CatIBs. The deviation of the values for the last time point from the overall trends is explained by artifacts due to high cell densities.
  }
    \label{plots}
\end{figure}

The proposed automated HIPP, applied to the 15~images for each time point, was used to observe the CatIB formation process over the whole cultivation time along with the cell growth (Figure \ref{plots}). The first CatIBs were detected after 31.5~h. The amount of in-focus cells with 1~CatIB reached its maximal amount at about 60~\%, while by the end of the experiment the formation of the second CatIB and the CatIB area growth is still ongoing. The linear trend of cell area growth is noted, where cell area reaches its maximum (3~µm$^{2}$) by the end of the experiment. The growth of CatIB area over cell area fraction shows exponential behavior and reaches almost 20~\%.
Overall, the resulting numbers provide the necessary information for CatIB production characterization, being an enabler for image-based bioprocess control.%

\section{CONCLUSION}

In this paper, a fully automated microscopic screening system was established, saving 8 h of manual microscopy in one experiment. It was used to observe CatIB formation over the whole cultivation time. The system includes a hybrid approach combining ML-based detection with model-based unsupervised segmentation. Given the flexibility by choosing different segmentation parameters, and the adequately accurate results of the approach, despite low-quality images, we expect a similar strategy to be effective for comparable image segmentation tasks in microbiology and medicine, where no sufficient amount of training data is available to perform a fully data-driven segmentation approach.

\addtolength{\textheight}{-12cm}   %

\bibliographystyle{IEEEbib}
\bibliography{refs}

\end{document}